\documentclass[journal]{IEEEtran}

\usepackage{makecell}
\usepackage{float}
\usepackage{caption}
\usepackage{subcaption}
\usepackage{longtable}
\usepackage{tabularx}
\usepackage{tabu}
\usepackage{multirow}
\usepackage{tikz}
\usepackage{graphicx}
\usepackage{tikzscale}
\usepackage{pgfplots}
\usepackage{setspace}
\usepackage{booktabs}
\usepackage{siunitx}
\usepackage{adjustbox}
\usepackage{soul, color}
\usepackage{xcolor}

\usepackage[most]{tcolorbox}
\newtcolorbox{highlighted}{colback=yellow,coltext=red,breakable}

\usepackage{algorithmic}
\usepackage{amsmath,amssymb,amsfonts}
\usepackage[ruled,linesnumbered]{algorithm2e}
\usepackage[font=small,labelfont=bf]{caption}
\SetKwFor{Foreach}{for each}{do}{endForEach}%
\SetKwFor{For}{for}{do}{endfor for}%
\SetKwIF{uIf}{ElseIf}{Else}{if}{then}{else if}{else}{endIf if}%
\SetKwFor{While}{while}{do}{endWhile while}%

\begin{document}

\title{
Blockchain-based Federated Learning with Secure Aggregation in Trusted Execution Environment for Internet-of-Things
}

\author{Aditya~Pribadi~Kalapaaking, Ibrahim~Khalil, Mohammad~Saidur~Rahman, Mohammed Atiquzzaman, Xun~Yi, and Mahathir~Almashor}

% make the title area
\maketitle

\begin{abstract}
This paper proposes a blockchain-based Federated Learning (FL) framework with Intel Software Guard Extension (SGX)-based Trusted Execution Environment (TEE) to securely aggregate local models in Industrial Internet-of-Things (IIoTs). In FL, local models can be tampered with by attackers. Hence, a global model generated from the tampered local models can be erroneous. Therefore, the proposed framework leverages a blockchain network for secure model aggregation. Each blockchain node hosts an SGX-enabled processor that securely performs the FL-based aggregation tasks to generate a global model. Blockchain nodes can verify the authenticity of the aggregated model, run a blockchain consensus mechanism to ensure the integrity of the model, and add it to the distributed ledger for tamper-proof storage. Each cluster can obtain the aggregated model from the blockchain and verify its integrity before using it. We conducted several experiments with different CNN models and datasets to evaluate the performance of the proposed framework.
\end{abstract}

\begin{IEEEkeywords}
Federated Learning, Internet-of-Things, Blockchain, Secure Aggregation, Intel SGX, Trusted Execution Environment, Deep Learning 
\end{IEEEkeywords}

\section{Introduction}

The Internet-of-Things (IoT) explosion has made it an integral component of various intelligent applications. Intelligent applications include but are not limited to healthcare, manufacturing, critical system infrastructure, agriculture, and transportation. IoT devices enable the collection of a large volume of data and act autonomously in an intelligent system, thanks to machine learning algorithms. The large volume of IoT data plays an essential role in training a machine learning algorithm system. In general, IoT devices are resource-constrained and cannot execute machine learning algorithms independently. Edge computing technology is gaining acceptance at a tremendous rate to form intelligent networks in conjunction with IoT and machine learning. An edge device (referred to as an edge server throughout the article) and IoT devices within the network form a cluster. In an intelligent system, edge devices can host a machine learning algorithm that uses a locally-built dataset and produce a trained model. IoT devices generate data and receive control instructions depending on the type of IoT device. Later, the trained model can be used to make an intelligent decision in the system.

\begin{figure}[!h]
\centering
\includegraphics[width=0.83\linewidth]{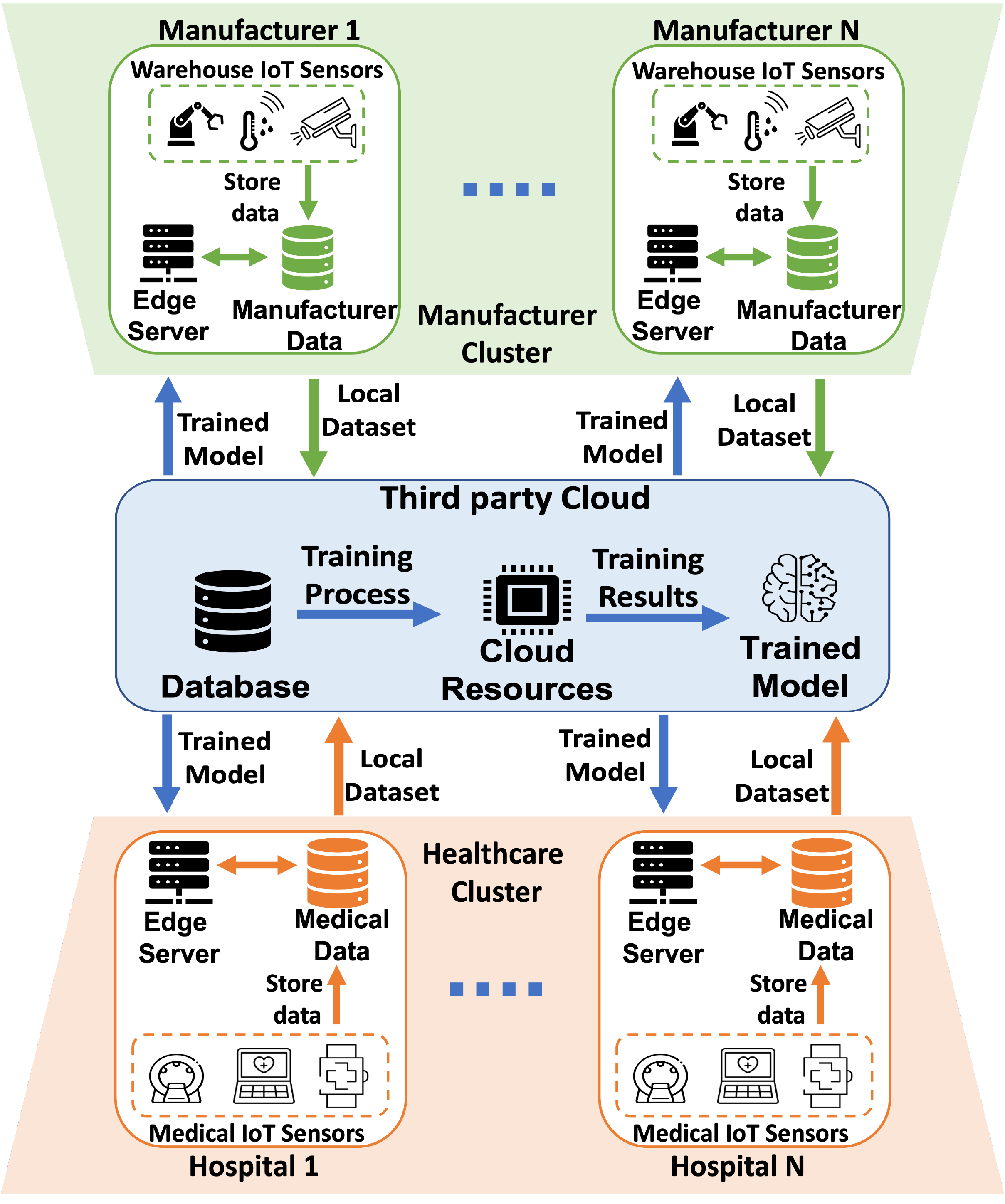}
\caption{Traditional collaborative learning application scenario}
\label{fig:scenario}
\end{figure}

Although an edge and IoT-based system configuration with
machine learning capability can manage different system tasks automatically, the level of accuracy impedes its success. For example, a trained model produced by an edge server with local data might not consider many features that could be absent in the local dataset. The accuracy can be improved if the edge device can collaborate with other edge servers that have produced their trained model based on their local datasets. This learning method is called \textit{Distributed Collaborative Machine Learning}\cite{alazab2021federated}. Traditional distributed collaborative machine learning (see Fig. \ref{fig:scenario}) allows different clusters to send their locally-trained model and datasets to a centralized server, such as the cloud. Cloud aggregates all locally-trained models using datasets from different sources and produces an aggregated trained model shared with all clusters to improve decision-making accuracy.

Distributed collaborative learning suffers from two significant issues: privacy and trust \cite{melis2019exploiting}. A new form of distributed collaborative learning, called \textit{Federated Learning (FL)} \cite{li2020federated}, enables different clusters to build a trained model with their local data, called a \textit{local model}, and to share only the local model with other participants for the purpose of aggregation. The aggregated model is known as a \textit{global model}. Data privacy is ensured because the global model is generated without the data being shared with other participants. Nevertheless, the global model cannot be fully trusted as internal or external attackers can launch several security attacks during the model aggregation and the dissemination of the global model. Hence, a trustworthy framework is required to ensure the privacy of sensitive data and the trustworthiness of the generated global model. Moreover, the receiver of the global model (e.g., edge server) should verify the integrity of the global model before using it.

\vspace{-5mm}\subsection{Contributions}
In this paper, we propose a complete framework for FL that simultaneously safeguards the privacy of IoT data and ensures security during the generation of aggregated trained models. In addition, the proposed framework guarantees trustworthy storage and sharing of the outcomes of any training. The proposed framework comprises a Convolutional Neural FL architecture that combines an Intel Software Guard Extension (SGX)-based Trusted Execution Environment (TEE) and blockchain platform. We assume that multiple IoT and edge devices clusters produce locally-trained models based on their local dataset and send the local model to the blockchain network for aggregation. In this framework, each blockchain node hosts an SGX-enabled processor that individually performs the FL- based aggregation tasks to generate an aggregated model. Once SGX-enabled processors of blockchain nodes perform the aggregation, each node can verify the authenticity of the aggregated model, run a blockchain consensus mechanism to ensure the integrity of the model, and add it to a blockchain for tamper-proof storage. An edge server from each cluster can collect the latest aggregated model from the blockchain and verify its integrity before using it. The key contributions of our work are summarized below. 
\begin{itemize}
    \item {The proposed framework introduces a new FL architecture for IoT to ensure secure generation of the aggregated model using Intel SGX-powered TEE.}
    \item {We propose the hosting of an SGX processor by a blockchain node that is responsible for the FL model aggregation task.}
    \item {A blockchain-powered trustworthy aggregated model storage and sharing model is proposed for FL-based learning in IoT applications.
    }
\end{itemize}

\vspace{-5mm}\subsection{Organization}
The rest of the paper is organized as follows. 
Section \ref{sec:issue} describe the problem scenario in collaborative learning. Section \ref{sec:related} discuss some of the closely related work. The proposed framework is described in Section \ref{sec:framework}. Section \ref{sec:exp} presents the experimental results and evaluates various performance aspects of the proposed framework. Section \ref{sec:con} concludes the paper.

\section{Problem Scenario}\label{sec:issue}

\begin{figure}[t]
\centering
\includegraphics[width=1\linewidth]{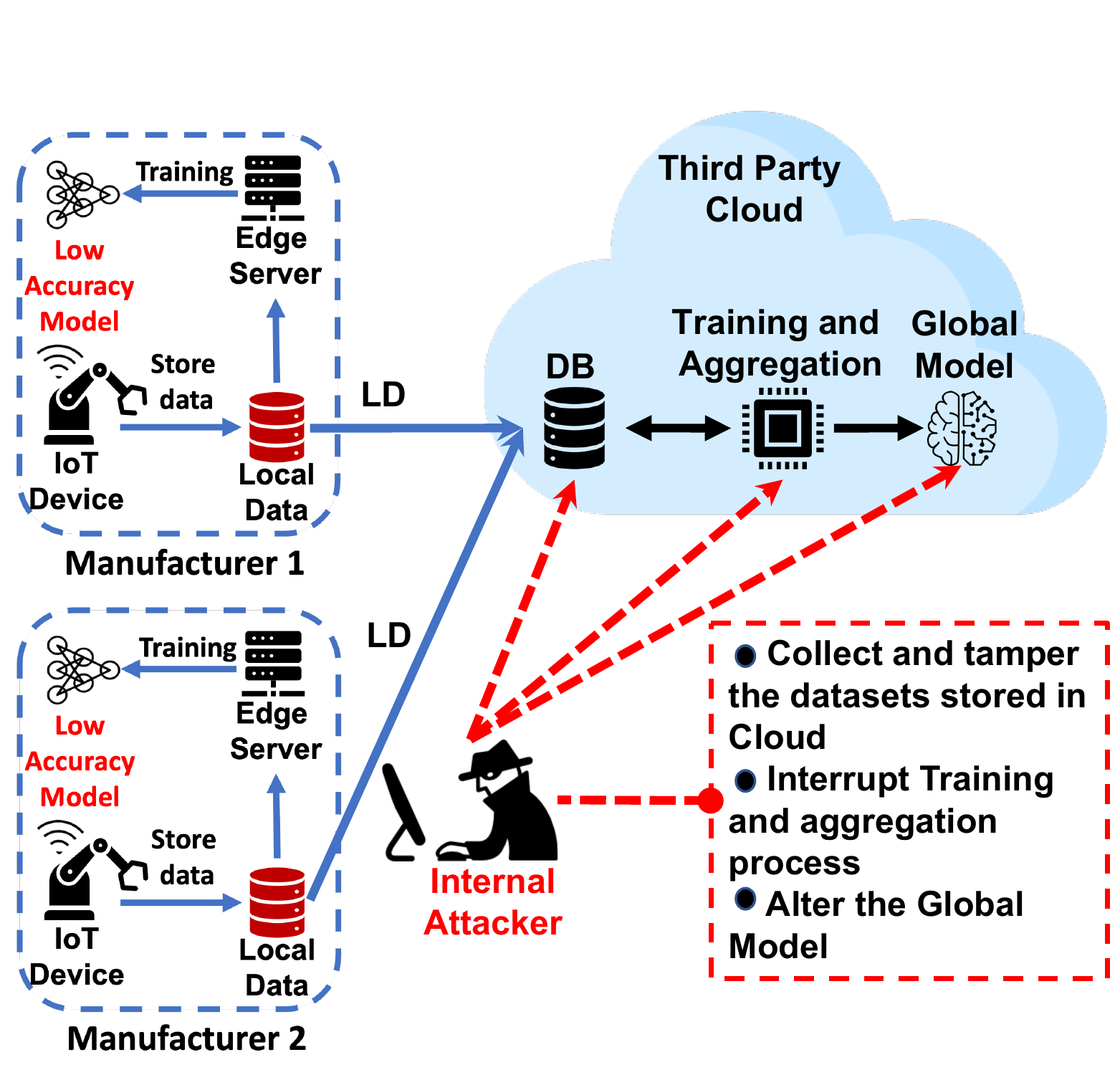}
\caption{Possible threat on collaborative learning architecture}
\label{fig:problem}
\end{figure}

To demonstrate and discuss the problem that exists with traditional collaborative machine learning, we use an IoT-enabled smart warehouse scenario (see Fig. \ref{fig:problem}). Assume that several smart warehouses are geographically dispersed. Each warehouse receives multiple pre-packed boxes of various garments (for both men and women), including shirts, trousers, shoes, jackets, and bags for storage. Each warehouse uses machine learning and an IoT-enabled camera to automatically sort the boxes according to the type of garment they contain. The camera scans the generic photo of the garment, which is shown on the box. However, IoT-enabled cameras are resourced-constrained and cannot execute the machine learning algorithm. Hence, each warehouse is equipped with an edge server with access to the local dataset and hosts the machine learning algorithm to train a model for recognizing garment items based on the local dataset. Nevertheless, the accuracy of a training model derived from the local dataset may not be good. Therefore, the edge server of each warehouse participates in a cloud-based collaborative machine learning platform to share its local dataset and the trained model. The cloud-based collaborative machine learning platform produces an aggregated model based on the received local datasets and models. The aggregated model is sent to all edge servers to achieve higher accuracy in recognizing the garment items.

Although the aforementioned collaborative learning scenario improves overall accuracy, it suffers from the following security risks:
\begin{itemize}
    \item {\textit{Risks of data privacy:} 
Sending local datasets to the cloud introduces the risk of a privacy breach. For example, a dishonest employee from the cloud service provider can act as an \textit{internal attacker} and collect the warehouse’s sensitive product information and share it with a business competitor for financial gain. Hence, there is the need for an aggregation model that would not require local datasets to generate an aggregated model.
}
    \item {\textit{Risks of generating biased aggregated trained model:} 
The aggregated model produced by a cloud service provider can be biased, as a cloud-based platform cannot be trusted. For instance, an internal attacker can generate a biased aggregated model not using the given local models or inject a faulty trained model to interrupt the generation of aggregated models. Therefore, a secure environment is required to prevent biased model generation.
}
    \item {\textit{Risks of receiving alteration or faulty aggregated trained model:} 
In the traditional cloud-based collaborative learning environment, an internal attacker of the cloud platform can interfere with disseminating the aggregated model. For example, an attacker can alter some part of the aggregated model before the cloud sends it to the edge servers. The traditional method does not allow a receiver of the aggregated model (i.e., edge server) to verify its integrity before using it. Hence, a trustworthy platform is required for sharing the aggregated model with edge servers.
    }
\end{itemize}

\section{Related Work}\label{sec:related}
This section discusses several studies that are closely related to our work.

\textbf{Privacy-preserving Federated Learning.} 
Several works on privacy-preserving federated learning have been presented recently. Yin \textit{et al.} \cite{yin2021privacy} and Liu \textit{et al.} \cite{9184079} proposed a federated learning framework where the training is performed on each node and only the model is sent to the central server to perform the model aggregation. Wei \textit{et al.} \cite{wei2020federated} and Zhao \textit{et al.} \cite{9253545} proposed a framework where data privacy is improved by means of differential privacy. However, the use of DP will slow down the training process and reduce accuracy. In \cite{zhao2021anonymous} the author proposes anonymous federated learning by adding a proxy layer and DP to the data. However, the proxy layer will add communication overhead, and the result shows that the DP decreases the ML accuracy. Li \textit{et al.} \cite{9187932} leverage SMPC-based federated learning to secure aggregation. Hence, their framework relies on a centralized server to arrange the secret sharing. This could be a problem since all the models can be seen in plaintext after the cloud collects the secret share. Federated learning is delicate to an attacker that can launch backdoor attacks. Bagdasaryan \textit{et al.} \cite{bagdasaryan2020backdoor} found that a backdoor can compromise the federated learning and poison the machine learning model. Our framework will create a secure end-to-end federated learning process to overcome this problem by securing the machine learning model and the aggregation process.

\textbf{TEE-based Machine Learning.}
Recently, TEE has gained popularity in the field of privacy-preserving machine learning. Ohrimenko \textit{et al.} \cite{ohrimenko2016oblivious} investigated centralized machine learning processes in an SGX-enabled data center to improve data privacy and avoid data leaks. In his framework, the server requests the dataset from all the participants and computes it in a centralized server. Tramer \textit{et al.} \cite{tramer2018slalom} and Juvekar \textit{et al.} \cite{juvekar2018gazelle} proposed a secure inference process inside of the TEE. Hynes \textit{et al.} \cite{hynes2018efficient} and Hunt \textit{et al.} \cite{hunt2018chiron} demonstrated centralized privacy-preserving machine learning by running all the CNN processes inside the enclave. 

The available frameworks use a single deep learning model, and none of them performs within the federated learning setup. The current work also shows that the time cost is significantly increased when the training process is performed in the TEE. Hence, we run the aggregation process inside the enclave to maximize the performance and reduce time consumption.

\textbf{Blockchain-based Federated Learning.}
Blockchain was first launched as a cryptocurrency technology. However, it has now been expanded for data storage across multiple computational nodes in a distributed fashion. Blockchain is structured as a linked list of blocks holding a set of transactions. Ali \textit{et al.} \cite{ali2021security} proposed a method to ensure the privacy and security of healthcare systems using blockchain. Their approach focuses mainly on securing patient data from active collision attacks by leveraging novel smart contracts and encryption algorithms. Nowadays, many studies are incorporating blockchain into their federated learning methodologies because federated learning is based on a centralized server, which is vulnerable to attack. Zhao \textit{et al.} \cite{9170559} designed a system where each of the clients will sign the model after the training process and send it to the blockchain. However, if this model has many clients, the computation cost will be very high. In recent works, \cite{9134967,9164912}, and \cite{feng2021blockchain} proposed a framework where the model is stored in the blockchain node, and federated learning is performed. However, in their architecture, the model is not totally encrypted. Also, the aggregation is performed by an untrusted party. Kim \textit{et al.} \cite{8733825} proposed a method where they deploy the blockchain on the edge devices. The disadvantage of this method is that the edge devices will require a lot of computation power. The author in \cite{kumar2021blockchain} proposes a blockchain architecture to collect the locally-trained model weights collaboratively from different sources for healthcare scenarios. However, the local model that is stored in the blockchain is not protected by any privacy measure. In this case, other parties can see the model, thereby raising privacy issues.

Samuel \textit{et al.} \cite{9684698} proposed blockchain-based FL for healthcare system. Their proposed framework protects the local model training with differential privacy (DP). The central server aggregates the global model and stores it in the blockchain. However, the global model accuracy is lower than the locally trained model. The use of DP in this framework can preserve privacy while sacrificing accuracy. Alsamhi \textit{et al.} \cite{9635590} and Otoum \textit{et al.} \cite{9670460} proposed an edge intelligence over smart environments with the support of FL and blockchain. Their proposed architectures leverage drones as an edge intelligence to perform the aggregation in FL. The aggregation process on a drone is vulnerable to tampering attacks and poisoning attacks. Since drones are deployed on the field and open networks, hardware security such as TEE can secure the aggregation process.

In Table \ref{summary}, we summarize some of the works to identify their research gaps and discuss how our proposed method differs from them. As shown in the table, existing works are mostly unsecured, inefficient, and have low accuracy. Hence, we deploy the blockchain on the server-side to reduce training model storage costs and leverage TEE to ensure secure and trustworthy model aggregation before sending it to the blockchain.

%%%%
\begin{table}[h]
\begin{center}
\scalebox{0.7}{
{
    \begin{tabular}{ 
    |p{0.12\textwidth}
    |p{0.24\textwidth}
    |p{0.24\textwidth}| }
        \hline
        
        \textbf{Methodology} & \textbf{Description} & \textbf{Remarks}\\
        
        \hline
        Yin \textit{et al.} \cite{yin2021privacy}
        &A privacy-preserving machine learning approach based on DP and sparse vector technique.
        &Low model accuracy and security is poor.\\
        \hline
        Wei \textit{et al.} \cite{wei2020federated}
        &A FL approach with DP to secure the data on edge devices.
        &Centralized approach with lower accuracy and efficiency.\\
        \hline
        Ohrimenko \textit{et al.} \cite{ohrimenko2016oblivious}
        &TEE-based machine learning approach.
        &Centralized approach with lack of data privacy. TEE is leveraged during local training. Hence, may not be suitable for Edge devices.\\
        \hline
        Hunt \textit{et al.} \cite{hunt2018chiron}
        &TEE-based privacy-preserving Machine Learning approach
        &Secure model generation through TEE; however, the global model is generated and stored in the unsecure centralized server.\\
        \hline
        Qu \textit{et al.} \cite{qu2020decentralized}
        &A blockchain-based FL approach. Blockchain is used to store model securely. Edge for cognitive computing in industrial IoT. Their framework keeps the local data on their edge devices and uses blockchain to secure the model.
        &The model aggregation is performed in an untrusted environment; hence, susceptible to tampering attacks.\\
        \hline
        Kim \textit{et al.} \cite{8733825}
        &A blockchain of edge devices and FL based fully decentralized approach.
        &Inefficient and the blockchain-based design is too heavy to implement in edge devices.\\
        \hline
    \end{tabular}
    }
    }
    \end{center}
    \caption{Summary of Related Works}
    \label{summary}
\end{table}
%%%%

\section{Proposed Framework}\label{sec:framework}
In this section, we present the proposed blockchain-based federated learning with Trusted Execution Environment (TEE)-powered secure aggregation framework. First, we present an overview of the system architecture. Next, we discuss in detail the various components of our proposed framework.

\begin{table}
\begin{center}
\caption{Notations}
\scalebox{0.7}{
\begin{tabularx}{\linewidth}{@{}XX@{}}
\toprule
    $M_{L}$ & Local Model \\
    $M_{G}$ & Global Model \\
    $M_{Li}^{r+1}$ & Updated Local Model\\
    $M_{Gi}^{r+1}$ & Updated Global Model\\
    $D_{i}$ & Local Image Dataset\\
    $C_{i}$ & IoT Cluster\\
    $E_{i}$ & Edge Server\\
    $B_{i}$ & Blockchain Node\\
    $S_{i}$ & SGX-enabled CPU\\
    $E(M_{Li},K_i)$ & Encrypted Training model\\
    $R_{i}$ & Remote Attestation Report\\
    $Q$ & Quotation for Global Model\\
\bottomrule
\end{tabularx}
}
\end{center}
\end{table}

\subsection{System Architecture}
We consider a Federated Learning (FL)-based collaborative learning model in this system that leverages Trusted Execution Environment (TEE) for secure aggregation and blockchain for tamper-proof data sharing and storage.

\begin{figure}[!h]
\centering
\includegraphics[width=0.9\linewidth]{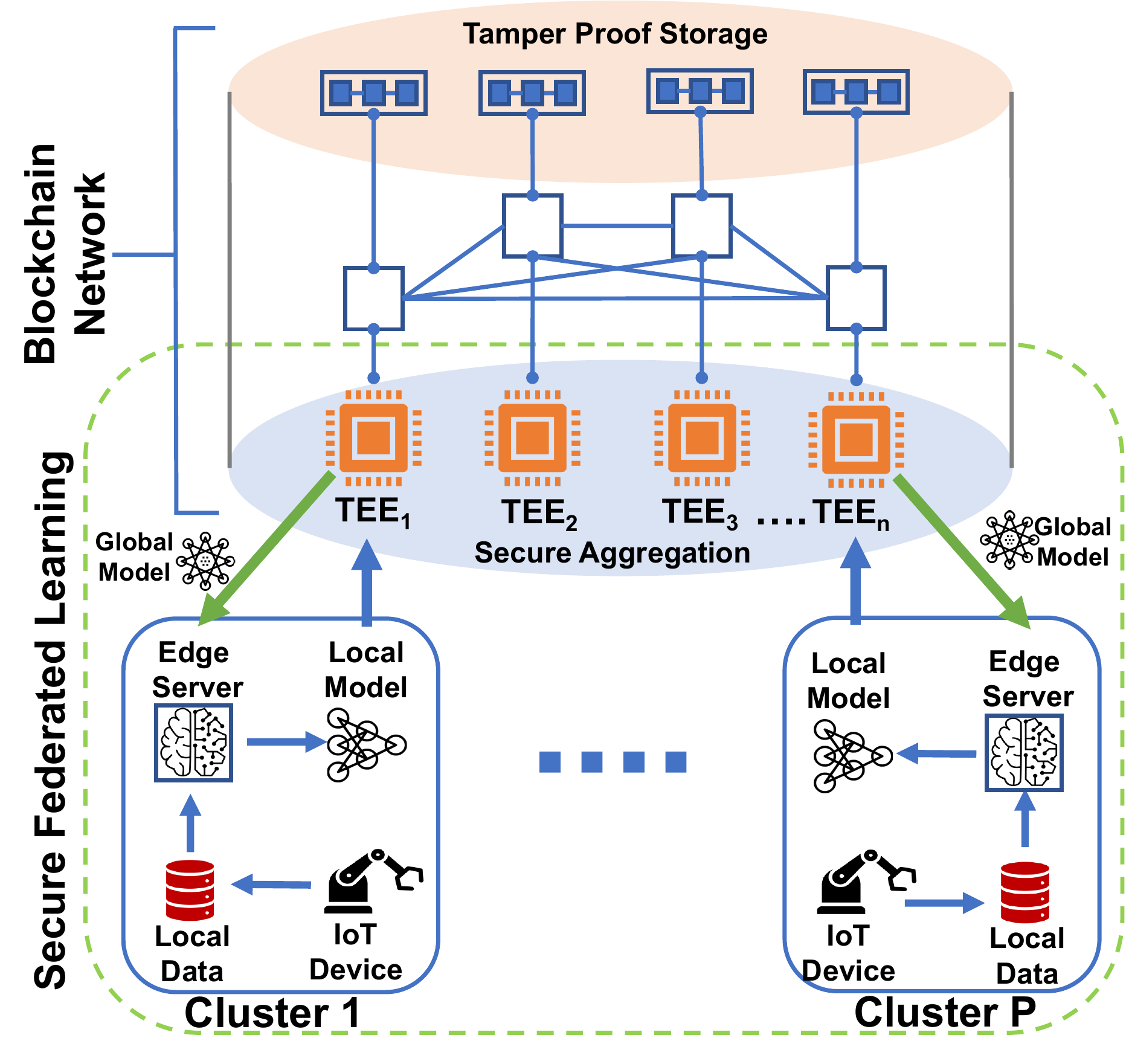}
\caption{Overview of the proposed framework}
\label{fig:architecture}
\end{figure}

We assume that there are $p$ warehouses equipped with several IoT cameras to scan product photos and recognize the type of products. Because IoT cameras are resource-constrained when running machine learning algorithms, each warehouse uses an edge server to host and execute a machine learning algorithm. As a result, IoT cameras and the edge server form a cluster $C_{i} (1 \leq i \leq p)$. Initially, the edge server trains a model based on the local dataset and generates a trained model called a \textit{Local Model}, denoted as $M_{L}$. However, if the size of the local dataset is small, the accuracy of $M_{L}$ might not be high. Hence, the edge server of a cluster $C_{i}$ joins in FL involving multiple clusters of similar warehouses by sending its $M_{L}$ to generate an aggregated model known as a \textit{Global Model}, denoted as $M_{G}$. In our proposed scenario, we adopt Federated Averaging (FedAVG) \cite{mcmahan2017communication} algorithm for generating the global model, which will be discussed in Section IV-C.

A typical FL approach involves three steps: \textit{initialization}, \textit{aggregation}, and \textit{update}. Unlike the traditional FL approach where $M_{L}$ are aggregated in a centralized server (e.g., a cloud server), our proposed framework uses a blockchain platform for the aggregation of $M_{L}$. Multiple nodes form a blockchain network, and each node receives all $M_{L}$ and individually aggregates $M_{L}$ to produce their own copy of a $M_{G}$. We assume that each blockchain node has a \textit{TEE host}. To ensure the security during the aggregation process, each blockchain node performs the aggregation in its TEE host and produces a $M_{G}$. Blockchain nodes execute a consensus mechanism to ensure that all nodes have identical $M_{G}$. Once the consensus has been reached, each blockchain node stores the $M_{G}$ in its respective blockchain. Finally, the blockchain network sends $M_{G}$ to all edge servers. Edge servers validate $M_{G}$ once received and update their initial model with $M_{G}$. Edge servers use the $M_{G}$ for product recognition in the warehouse.

Fig. \ref{fig:architecture} gives an overview of the proposed framework, which consists of three main phases: \textit{Local Model Generation}, \textit{Secure TEE-Enabled Aggregation}, and \textit{Blockchain-Based Global Model Storage}. The following subsections describe each phase in detail.

\vspace{-3mm}\subsection{Local Model Generation}
\begin{figure}[t]
\centering
\includegraphics[width=1.0\linewidth]{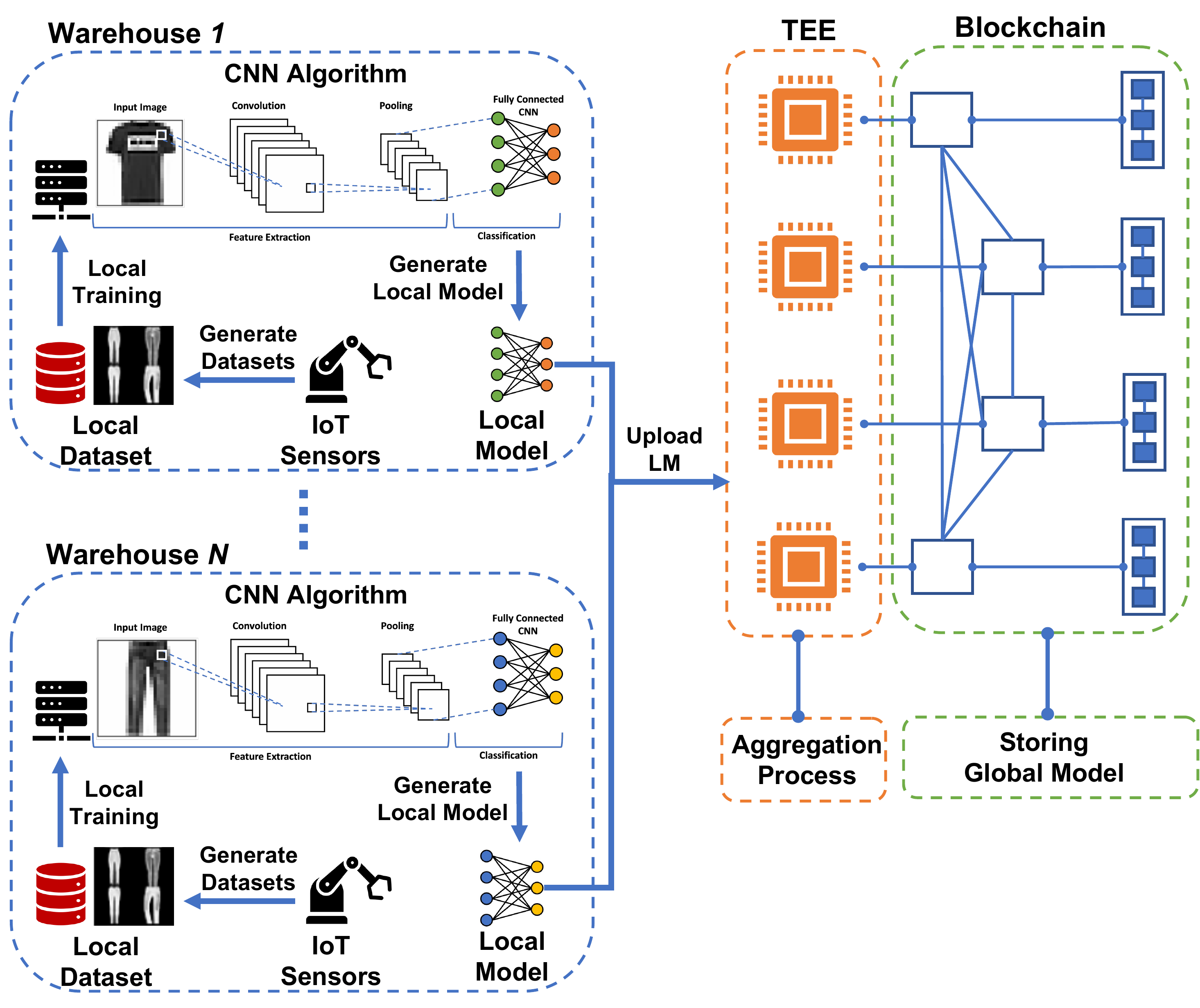}
\caption{Local model generation}
\label{fig:local_model_gen}
\end{figure}

The Local Model Generation (LMG) phase is performed in each cluster to generate a locally-trained model similar to the initialization phase of the original FL. An overview of the LMG phase is given in Fig. \ref{fig:local_model_gen}. In the proposed system, we assume that the edge servers of different clusters train models using Convolutional Neural Network (CNN)-based image classification in which model parameters are retrieved from the global model stored in the tamper-proof storage. Example of the CNN models are AlexNet\cite{krizhevsky2017imagenet}, LeNet\cite{lecun1998gradient} and VGG16\cite{simonyan2014very}.

In general, CNN image classification takes an input image, processes it and classifies it under certain categories of $t$ objects. An edge server $E_{i}$ of cluster $C_{i}$ has a local image dataset $D_{i}$.  The edge server sees an input image as an array of pixels, and it depends on the image resolution. Based on the image resolution, it will see $h \times w \times d$ ($h$ = Height, $w$ = Width, $d$ = Dimension). For example, an image of 6 x 6 x 3 array of a matrix of RGB (3 refers to RGB values) and an image of 4 x 4 x 1 array of a matrix of a grayscale image. Technically, the deep learning CNN model works via different layers to train and test a local model. The layers are \textit{convolution layers with filters (kernels)}, \textit{pooling}, and \textit{fully connected layers (FC)}. In the end, CNN applies the \textit{SoftMax function} to classify an object according to probabilistic values between $0$ and $1$. In each edge server, $E_{i}$ locally trained the machine learning model $M_{Li}$. An edge server $E_{i}$ updates the ML model using its dataset in every FL round $r$ as follows:

\begin{equation}
  M_{Li}^{r+1} = M_G^r - \eta \nabla F(M_G^r,D^i) 
\end{equation}

where $M_{Li}^{r+1}$ denotes the updated local model of client $i$, $M_G^r$ is the current global model, $\eta$ is the local learning rate, $\nabla$ is used to refer to the derivative with respect to every parameter, and $F$ is the loss function. Later, $E_{i}$ send $M_{Li}^{r+1}$ to the blockchain network and aggregated iteratively into a joint global model $M_{G}$.

To ensure the security of the local model, $E_{i}$ leverages symmetric key encryption algorithm, such as Advanced Encryption Standard (AES), to encrypt $M_{Li}$ before sending it to the blockchain nodes. We assume that the AES secret key between $E_{i}$ and the blockchain network is established using a secure key establishment mechanism, such as Diffie–Hellman key exchange mechanism. We do not discuss this process in detail as we leverage the state-of-the-art mechanism for encrypting the local model.

\subsection{TEE enabled Secure Model Aggregation}

Once different local models are received by a blockchain node, a TEE is used to securely aggregate all models. For TEE, we use Intel Software Guard Extension (SGX) \cite{costan2016intel} in this framework. SGX is a set of CPU extensions, which can provide isolated execution environments, named \textit{enclaves}, to protect the confidentiality and integrity of the data against all other software, even a compromised OS, on the platform. When a platform is equipped with an SGX-enabled CPU (such as a blockchain node in our framework), as an enclave, the memory, BIOS, I/O, and even power are treated as potentially untrustworthy. Firstly, the encrypted data is transmitted into an enclave for decryption. Then the decrypted data will be the input of function $f$. Finally, the output of $f$ will be encrypted and then sent to the outside of the enclave.

Using the same principle, FL's local model aggregation task is performed in the SGX-enabled CPU. We assume that there are $b$ blockchain nodes in the blockchain network, and each blockchain node $B_i (1\leq i \leq b)$ in the blockchain network is equipped with an SGX-enabled CPU $S_{i}$. A blockchain node $B_{i}$ cannot access the code and data within its SGX-enabled CPU $S_{i}$. 

Assume that a blockchain node $B_{i}$ receives the set $M_{L}$ of local models from all clusters which can be denoted as $M_{L} = \{M_{L1}, M_{L2}, \hdots, M_{Lp}\}$. $B_{i}$ sends $M_{Li}(1 \leq i \leq C_i)$ to $S_{i}$. The secure aggregation tasks of all local models in $M_{L}$ is done using multiple operations which are discussed below.

\subsubsection{Generation of Encrypted Local Models}
The SGX enclave receives only encrypted data to ensure security. Hence, $B_{i}$ needs to encrypt $M_{Li}$ before sending it to $S_{i}$. Let, $E(.,K)$ be a Symmetric Encryption (SE) algorithm $E(.,K_i)$ with a secret key $K_i$ that is shared between $B_{i}$ and its $S_{i}$. The shared secret key $K_i$ is established by leveraging a secure key exchange protocol such as Diffie-Hellman Key Exchange Protocol.

$B_{i}$ generates an encrypted local model $E(M_{Li},K_i)$.
$B_{i}$ sends $E(M_{Li},K_i)$ to $S_{i}$.

\begin{figure}[!h]
\centering
\includegraphics[width=1.0\linewidth]{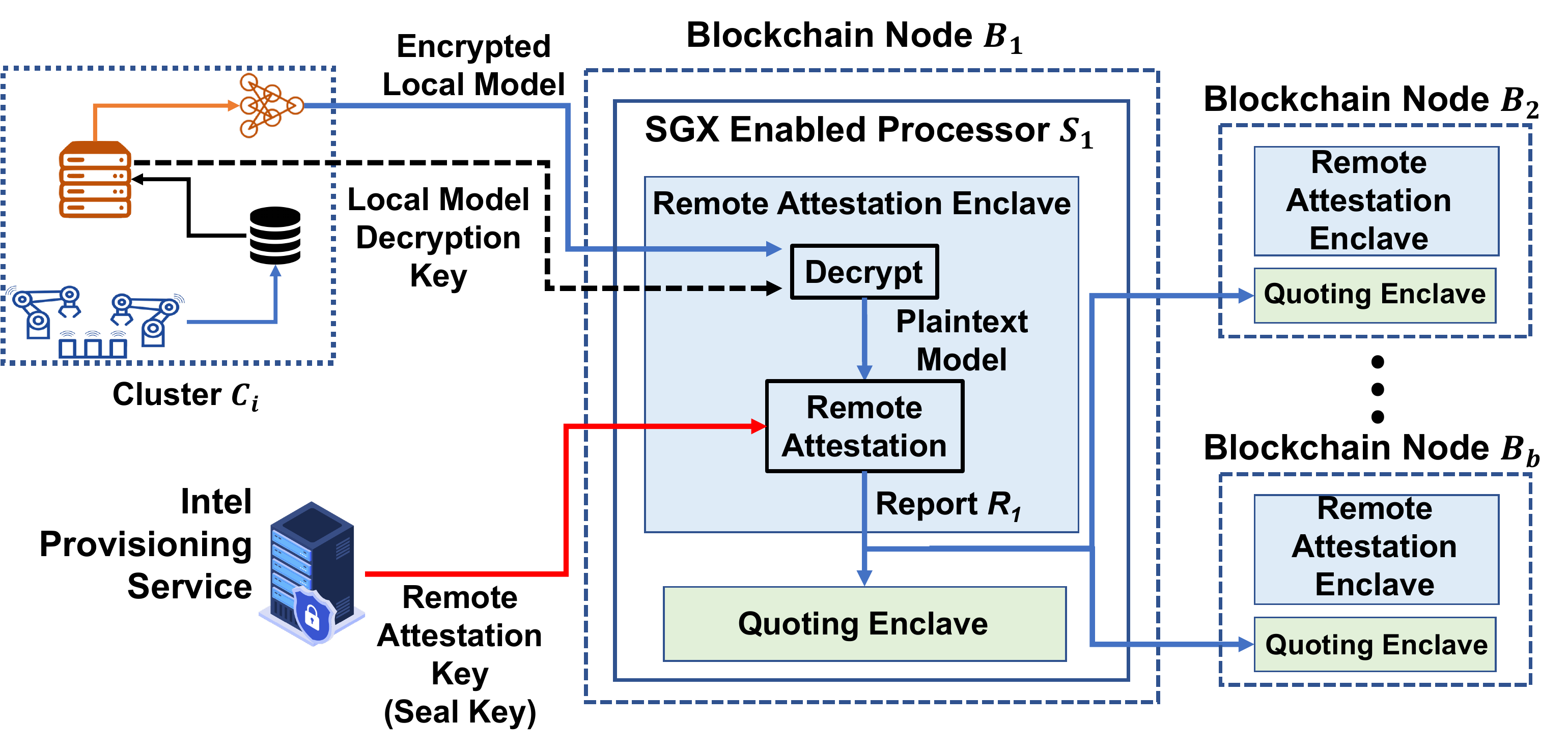}
\caption{Remote attestation of local model}
\label{fig:report}
\end{figure}

\subsubsection{Remote Attestation}
The remote attestation allows the verification of the integrity of the aggregated model (i.e., global model) generated by the SGX enclave. In this framework, the SGX enclave acts as the \textit{attestator}, and the software module of a blockchain node $B_{i}$ that is responsible for interfacing between SGX enclave and the blockchain network plays the role of a \textit{verifier} of the attestation. \textit{First}, the SGX enclave $S_{i}$ receives the set $M^{E}_{L}$ of $p$ encrypted local models which can be denoted as $M^{E}_{L} = \{E(M_{L1},K_1), \hdots, E(M_{Lp},K_p)\}$. $S_i$ decrypts each $E(M_{Li},K_i) (1 \leq i \leq p)$ with the shared secret key $k_i$ to retrieve the plaintext set of local learning models $M_{Li}$. \textit{Second}, $S_{i}$ performs the aggregation using the Federated Averaging (FedAVG)  \cite{mcmahan2017communication} algorithm to generate the global model $M_{G}$ as follows:

\begin{equation}
  M_{Gi}^{r+1} = \sum_{i=1}^{n} \frac{|D_i|}{N} M_{Li}^{r+1}, N = \sum_{i=1}^{n} |D_i|  
\end{equation}

where $M_{Gi}^{r+1}$ denotes the updated global model, $n$ is a number of clients on the federated learning round $r$, $|D_i|$ is the number of data items (images) owned by $E_{i}$ to train local model $M_{Li}^{r+1}$, and $N$ the total number of data used to train all of the local models. $M_{G}$ is final updated global model $M_{Gi}^{r+1}$.

\textit{Third}, $S_{i}$ generates a remote attestation, called \textit{report} $R_{i} = Sign(M_{Gi},A_{Ki})$. Here, $Sign(.,A_{Ki})$ is a signature function and $A_{Ki}$ is the attestation key of $S_{i}$. The generated report enables a verifier (i.e., blockchain node) to verify the $M_{Gi}$. The pseudocode of the overall aggregation and remote attestation is illustrated in Algorithm \ref{alg:aggregation}. The algorithm takes encrypted trained models as input and outputs aggregated global models, and its remote attestation report All tasks of Algorithm \ref{alg:aggregation} are executed under a running enclave of the SGX. SGX uses a quoting enclave to verify reports produced by the application enclave and signs as a quote. The quoting enclave is used to determine the trustworthiness of the platform. Later, the quote is sent to another party for verification. In our scenario, each $B_{i}$ will have one $S_{i}$ and works as an aggregator and verifier of attestation reports. Fig. \ref{fig:report} shows the details of the quoting enclave process.

\SetAlFnt{\small}
\begin{algorithm}
\caption{Aggregation Process and Remote Attestation in TEE}
\label{alg:aggregation}
\KwIn
{
    \begin{minipage}[t]{10cm}%
     \strut
      Encrypted Trained Models $M^E_{L} = \{E(M_{L1},K_1), \hdots, \\ E(M_{Lp},K_p)\}$
     \strut
  \end{minipage}%
}

\KwOut
{
    \begin{minipage}[t]{10cm}%
     \strut
     Aggregated Global Model ($M_{G}$) and \\
     Remote Attestation Report $R_{i}$
     \strut
  \end{minipage}%
}

\While{SGXServerRunning}
{
    \While{EnclaveRunning}
    {
        \textbf{Initialize:} \\
        Memory Buffer, $Mem = \emptyset$\\
        \Foreach{$E(M_{Li},K_i) \in M^E_{L}$}
        {
        Decrypt $E(M_{Li},K_i)$ with the key $K_i$ to obtain the corresponding decrypted local model $\overline{M_{Li}}$.\\
        Add $\overline{M_{Li}}$ to memory buffer $Mem$.\\
        }
        Aggregate all decrypted local models in $Mem$ according to \textbf{FedAvg} algorithm as shown in (3) and generate global model $M_{G}$.\\
        Generate a remote attestation $R_i$ for $M_{G}$ by signing it with the attestation key $A_{Ki}$ of $S_i$.\\
        \textbf{return} $\{M_{G}, R_i\}$
    }
}
\end{algorithm}

\begin{figure}[t]
\centering
\includegraphics[width=1\linewidth]{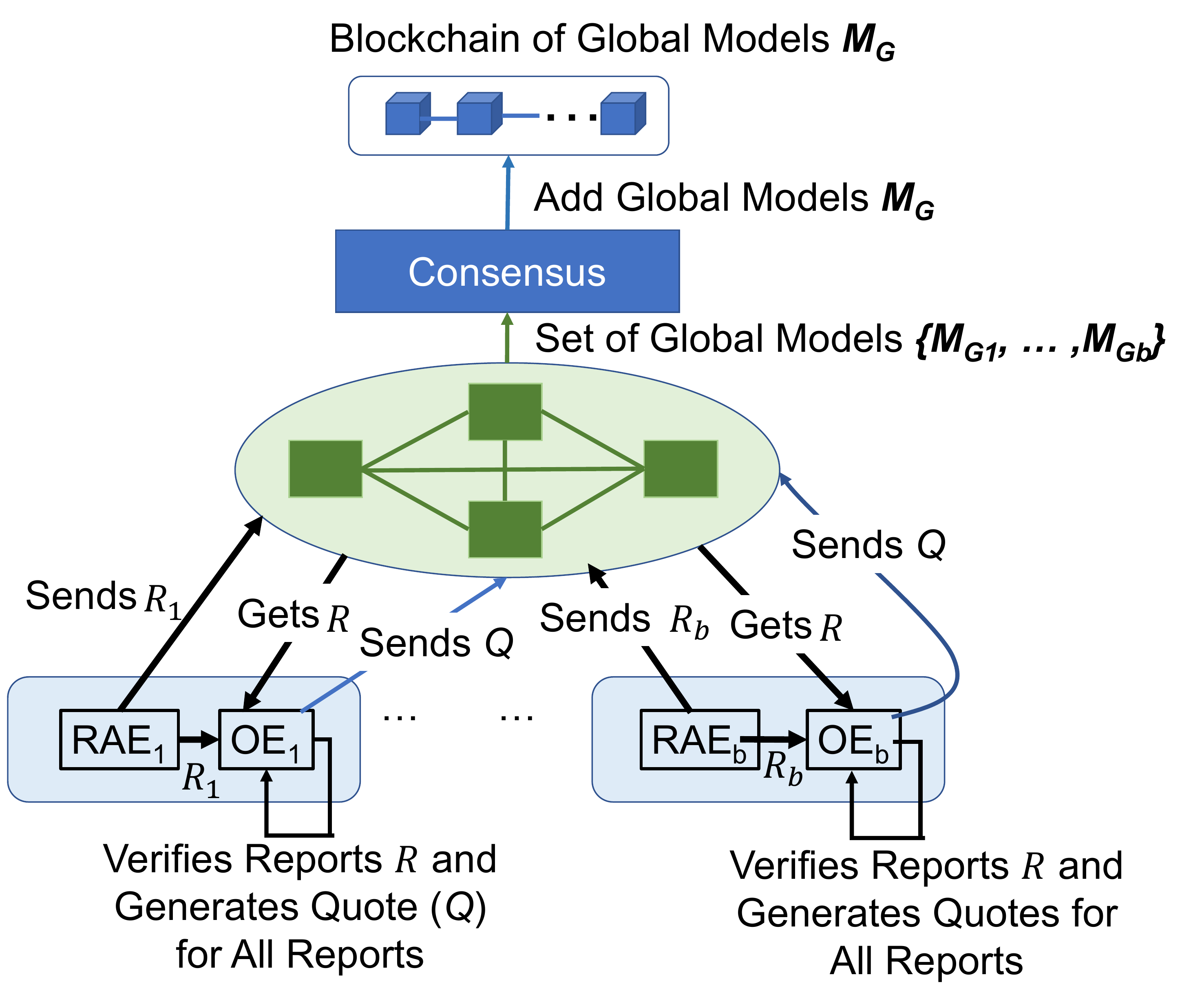}
\caption{Blockchain based global model storage}
\label{fig:storage}
\end{figure}

\subsection{Blockchain-based Tamperproof Global Model Storage and Distribution}
In this phase, the blockchain network receives all remote attestations produced by SGX enclaves and runs a consensus mechanism. The consensus mechanism verifies the remote attestations of a global model produced by the SGX enclaves. If all remote attestations are verified, and the majority hashes of corresponding models are the same, the blockchain nodes in the blockchain network add the global model $M_{G}$ as a block in the blockchain. Also, the global model is sent to all edge servers as the update operation FL. An overview of this step is given in Fig. \ref{fig:storage}.

\subsubsection{Verifying Attestation Reports by a Blockchain Node}
Assume that each blockchain node is equipped with a \textit{quoting enclave} and has an attestation key $A_{Kj}$ to sign a remote attestation report $R_i$ produced by $S_i$. $R_i$ is signed with $A_{kj}$ to generate a \textit{quote} $Q_i$. A quote contains the identity of the attesting enclave $S_i$, execution mode details (e.g. Security Version Number level $S_i$), and additional metadata. The function that is used to generate $Q_i$ can be shown as: $Q_{i} = Sign(R_{i},A_{Kj})$. $Q_i$ is encrypted using the public key $PK_{IAS}$ of Intel Attestation Service (IAS) and generates $E(Q_i, PK_{IAS})$. $PK_{IAS}$ is embedded in the quoting enclave of all SGX-enabled processors. Each $S_i$ shares its $E(Q_i, PK_{IAS})$ to other SGX-enabled processors of the blockchain network.
Once all encrypted quotes are received from SGX-enabled processors of all $n$ blockchain nodes, $B_{i}$ creates a collection of Encrypted Quotes received from all which is denoted by $Q^{E} = \{E(Q_1, PK_{IAS}), E(Q_2, PK_{IAS}), \hdots, E(Q_n, PK_{IAS})\}$.

A blockchain node $B_{i}$ verifies each encrypted quote with the help of $IAS$ and determines if the quote is correct and the corresponding remote attestation enclave has created it. The verification is done using a function $verify(E(Q_i, PK_{IAS}), PR_{IAS})$, where $PR_{IAS}$ is the private key of IAS. Once the quotation is verified, $Q$ is broadcast to the blockchain network to obtain a consensus for the global model. Algorithm \ref{alg:quote} provides an overview of this step.

\subsubsection{Consensus by Blockchain Network}
\begin{figure}[t]
\centering
\includegraphics[width=0.9\linewidth]{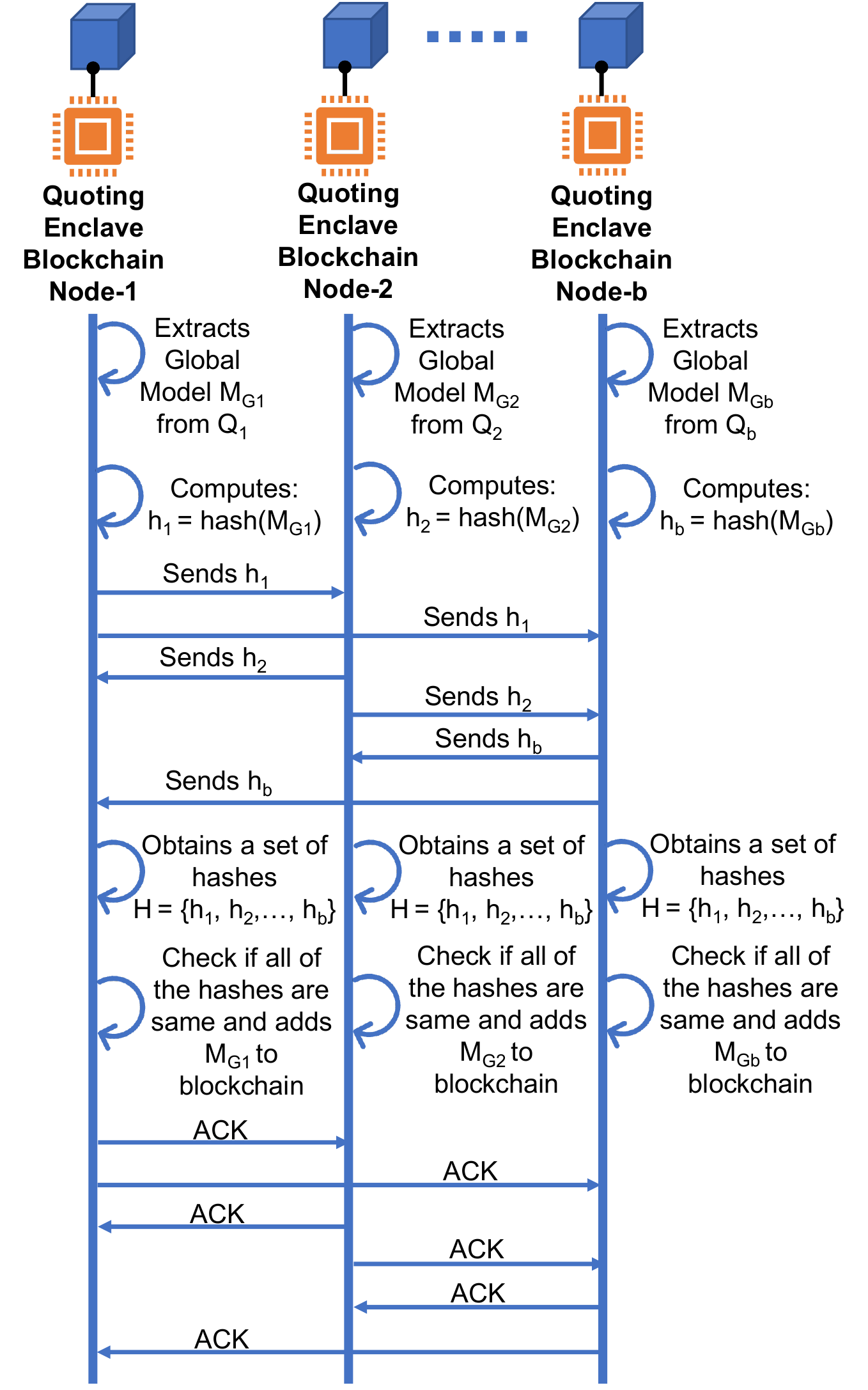}
\caption{Overview of the consensus mechanism}
\label{fig:consensus}
\end{figure}

\begin{figure}[t]
\centering
\includegraphics[width=1\linewidth]{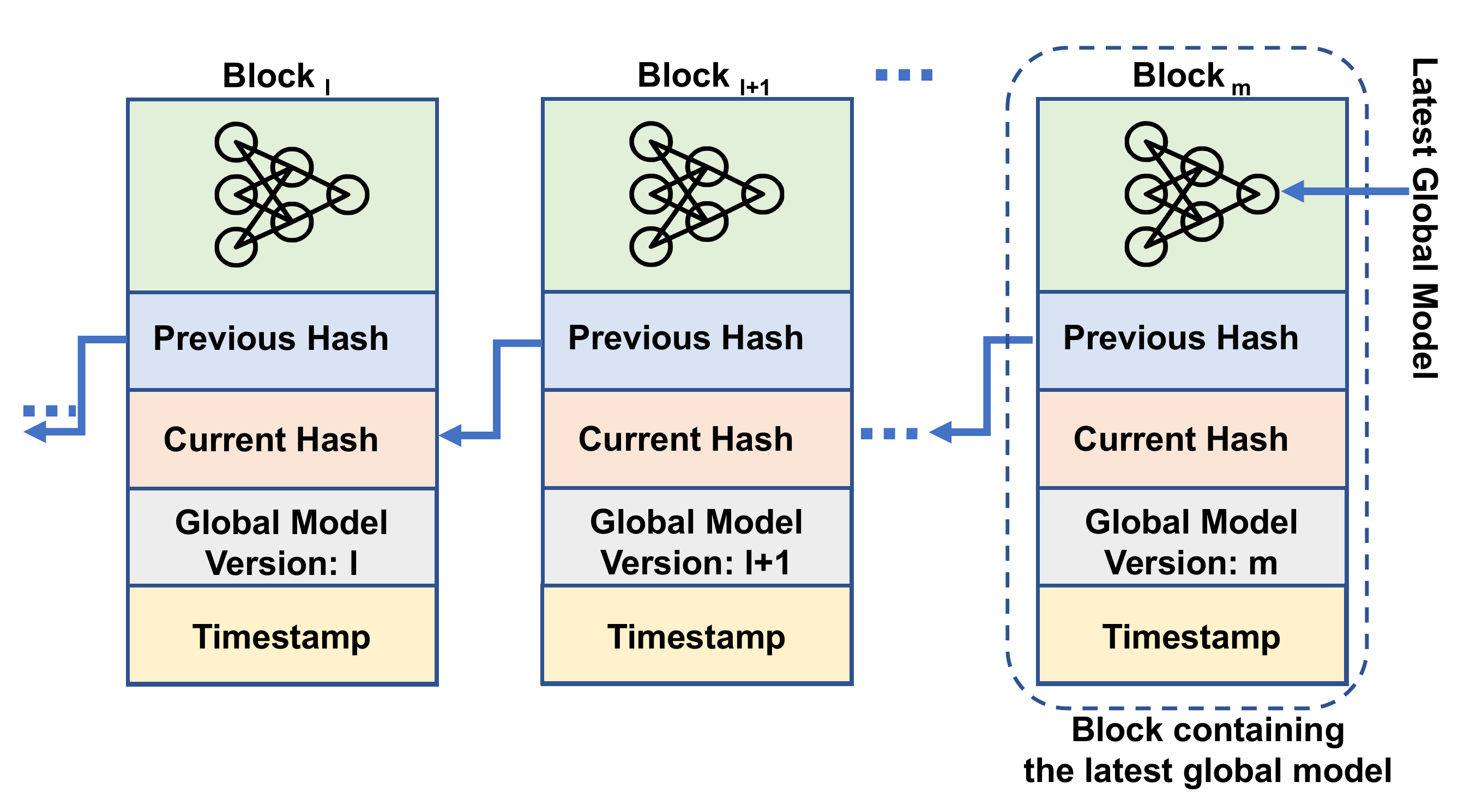}
\caption{Data structure of the global model blockchain}
\label{fig:data_structure}
\end{figure}

The consensus mechanism has several steps. \textit{First}, a blockchain node $B_{i}$ checks the validity of each quote and the authenticity of the quote-generating enclave. \textit{Second}, the global model is extracted from each quote, and their hashes are verified. If the hashes of all global models are the same, the consensus is achieved. If all hashes are not the same, the blockchain node $B_{i}$ determines the global model $M_{Gk}$ that has maximum matched hash values, where $k\leq p$. Third, $B_{i}$ proposes $M_{Gk}$ to the blockchain network to add in the blockchain. Finally, if $M_{Gk}$ is the same for the majority of the node's global model, the consensus is achieved and added to the blockchain. Algorithm \ref{alg:consensus} shows the pseudocode and Fig. \ref{fig:consensus} provide an overview of of consensus mechanism. The blockchain data structure of global models is illustrated in Fig. \ref{fig:data_structure}.

\begin{algorithm}
\caption{Quote Generation for Global Model}
\label{alg:quote}
\KwIn
{
    \begin{minipage}[t]{10cm}%
     \strut
      Remote Attestation Reports $R = \{(R_{1},Pk_1), \hdots, \\ (R_{p},Pk_p)\}$
     \strut
  \end{minipage}%
}

\KwOut
{
    \begin{minipage}[t]{10cm}%
     \strut
     Quotation for Global Model ($Q$) 
     \strut
  \end{minipage}%
}

\While{SGX Server is Running}
{
    \While{While Quoting Enclave is Running}
    {
        Collection of Quotes, $Q = NULL$\\
        \Foreach{$R_{i} \in R$}
        {
            Verify the validity of the report $R_{i}$ done by Enclave $S_{i}$\\
            Generate a Quote, $Q_{j} = Sign(R_{i},A_{kj})$\\
            $Q.add(Q_{j})$
        }
        Broadcasts $Q$ to Blockchain Network
    }
}
\end{algorithm}

\begin{algorithm}
\caption{Consensus Mechanism on Global Model}
\label{alg:consensus}
\KwIn
{
    \begin{minipage}[t]{10cm}%
     \strut
      Quotation for Global Model ($Q$)
     \strut
  \end{minipage}%
}

\KwOut
{
    \begin{minipage}[t]{10cm}%
     \strut
     Global Model ($M_{G}$)
     \strut
  \end{minipage}%
}

\While{Blockchain Node $B_{i}$ is Running}
{
    Set of Global Models, $GM = NULL$\\
    Set of Hashes of Global Models, $H = NULL$\\
        \Foreach{$Q_{j} \in Q$}
        {
            Verify the validity of the Quote $Q_{i}$ done by Enclave $S_{i}$\\
            Extract global model $M_{Gj}$ from Quote $Q_{j})$\\
            Add $M_{Gj}$ to $GM$ \\
            Send $M_{Gj}$ to other nodes in the blockchain network\\
        }
        \Foreach{$M_{Gj} \in GM$}
        {
            Get the hash $h_{j}$ of $M_{Gj}$ and it to $H$\\
        }
        Add $M_{Gj}$ to Blockchain if all hashes in $H$ are same.
}
\end{algorithm}

\section{Results and Discussion}\label{sec:exp}
In this section, we report on several experiments conducted to evaluate the performance of our proposed framework. Experimental setup, and dataset and model are discussed in Section \ref{sec:setup} and \ref{sec:data}, respectively. Section \ref{sec:results} shows experimental results and evaluates the performance.   

\vspace{-2mm}\subsection{Experimental Setup}\label{sec:setup}
In our experiments, both the server and participant applications were run on an Azure Cloud. We used the DCsv2 series VM with 4 vCPU and 16 GB Memory. This DC series from Azure provides confidentiality and integrity of the data and code while they are being processed in the public cloud. DCsv2-series using Intel® Software Guard Extensions was used, which enables the end-user to use secure enclaves for protection. These machines are backed by 3.7 GHz Intel® Xeon E-2288G (Coffee Lake) with SGX \cite{costan2016intel} technology. We built our federated learning application based on PyTorch \cite{paszke2019pytorch} and PySyft \cite{ziller2021pysyft}. To run the PyTorch application in the SGX environment, we build our application on GrapheneOS \cite{tsai2017graphene}. 

\subsection{Datasets and Model}\label{sec:data}

For the experiments, we selected three datasets popularly used for the machine learning process: Fashion MNIST \cite{xiao2017fashion}, CIFAR-10 \cite{krizhevsky2010convolutional}, and MNIST \cite{deng2012mnist}. These datasets are commonly used for benchmarking in the machine learning framework. Therefore, we have used them to evaluate the performance of our proposed approach. The dataset is used to train and test the local model on the client-side in the proposed FL-based approach. For all our experiments, we split the training and testing sets. Based on the number of participants, we evenly distribute the training and test sets among all participants. Fashion MNIST \cite{xiao2017fashion} is a collection of datasets containing fashion images. The training set comprised 60,000, and 10,000 images were used as a test set. Each image had a 28×28-pixel grayscale, and nine different classes were represented (trousers, dress, bag, etc.). MNIST \cite{deng2012mnist} is a dataset consisting of handwritten digits (60,000 images in the training set and 10,000 in the test set). Each image is a 28×28-pixel image of a handwritten digit. CIFAR-10 \cite{krizhevsky2010convolutional} consists of 50,000 images in the training set and 10,000 in the test set. It comprises 10 different classes (such as cars, dogs, planes), and there are 6,000 images in each class, where each image contains 32×32-colored pixels. Table \ref{tab3} shows the overview of the dataset used in the experiments.

We consider three models for our experiment. First, the LeNet model was used, which was proposed by LeCun et al. \cite{lecun1998gradient}. The model contains two convolutional layers and two fully-connected layers. This model is suitable for running experiments using the Fashion MNIST and MNIST datasets. Second, the AlexNet \cite{krizhevsky2017imagenet} model is used with five convolutional layers and three fully-connected layers. This model can use batch normalization layers for stability and efficient training. AlexNet is suitable for testing on the CIFAR-10 datasets. Finally, the VGG16 \cite{simonyan2014very} model is used that has 16 layers and about 138 million parameters. This machine learning model is also suitable for CIFAR 10 datasets.

%%%%
\begin{table}[h]
\begin{center}
\resizebox{\columnwidth}{!}{
    \begin{tabular}{ |c|c|c|c|c| }
        \hline
        \textbf{Datasets} & \textbf{Training set} & \textbf{Test set} & \textbf{Size} & \textbf{Color}\\
        
        \hline
        MNIST \cite{deng2012mnist} & 60.000 & 10.000 & 28x28 & Grayscale\\
        \hline
        F-MNIST \cite{xiao2017fashion} & 60.000 & 10.000 & 28x28 & Grayscale\\
        \hline
        CIFAR-10 \cite{krizhevsky2010convolutional} & 50.000 & 10.000 & 32x32 & RGB\\
        \hline
    \end{tabular}
    }
    \end{center}
    \caption{Datasets specifications}
    \label{tab3}
\end{table}
%%%%

\vspace{-2mm}\subsection{Experimental Results and Performance Evaluation}\label{sec:results}

%%%%
\begin{figure}[tbh!]
\centering
\begin{subfigure}[tbh!]{0.4\columnwidth}
    \resizebox{1\columnwidth}{!}
    {
        \begin{tikzpicture}
                \begin{axis}[
                    xlabel={Edge Devices},
                    ylabel={Processing Time (S)},
                    symbolic x coords = {2,10,20,30,40},
                    xticklabel style={anchor= east,rotate=45 },
                    xtick=data,
                    ymax=25,
                    ymin=0,
                    legend pos=north west,
                    ymajorgrids=true,
                    grid style=dashed,
                    legend style={nodes={scale=1, transform shape}},
                    label style={font=\Large},
                    tick label style={font=\Large}
                ]
                \addplot+[mark size=3pt]
                    coordinates {
                        (2,0.8)
                        (10,3.9)
                        (20,8.6)
                        (30,12.2)
                        (40,16.5)
                    };
                \addplot+[mark size=3pt]
                    coordinates {
                        (2, 0.8)
                        (10,4.5)
                        (20,9.4)
                        (30,14.8)
                        (40,19.3)
                    };
                \legend{Without SGX, Using SGX}
                \end{axis}
        \end{tikzpicture}
        }
    \caption{LeNet - Fashion MNIST}
    \label{exp1a}
\end{subfigure}
~
~
~
\begin{subfigure}[tbh!]{0.4\columnwidth}
    \resizebox{1\columnwidth}{!}{
    \begin{tikzpicture}
                \begin{axis}[
                    xlabel={Edge Devices},
                    ylabel={Processing Time (S)},
                    symbolic x coords = {2,10,20,30,40},
                    xticklabel style={anchor= east,rotate=45 },
                    xtick=data,
                    ymax=25,
                    ymin=0,
                    legend pos=north west,
                    ymajorgrids=true,
                    grid style=dashed,
                    legend style={nodes={scale=1, transform shape}},
                    label style={font=\Large},
                    tick label style={font=\Large}
                ]
                \addplot+[mark size=3pt]
                    coordinates {
                        (2,0.9)
                        (10,4.5)
                        (20,9.2)
                        (30,14.4)
                        (40,19.1)
                    };
                \addplot+[mark size=3pt]
                    coordinates {
                        (2,1)
                        (10,5.2)
                        (20,11.2)
                        (30,16.7)
                        (40,21.8)
                    };
                \legend{Without SGX, Using SGX}
                \end{axis}
            \end{tikzpicture}
        }
    \caption{VGG16 - CIFAR-10}
    \label{exp1b}
\end{subfigure}
~
~
~
\begin{subfigure}[tbh!]{0.4\columnwidth}
    \resizebox{1\columnwidth}{!}
    {
    \begin{tikzpicture}
                \begin{axis}[
                    xlabel={Edge Devices},
                    ylabel={Processing Time (S)},
                    symbolic x coords = {2,10,20,30,40},
                    xticklabel style={anchor= east,rotate=45 },
                    xtick=data,
                    ymax=25,
                    ymin=0,
                    legend pos=north west,
                    ymajorgrids=true,
                    grid style=dashed,
                    legend style={nodes={scale=1, transform shape}},
                    label style={font=\Large},
                    tick label style={font=\Large}
                ]
                \addplot+[mark size=3pt]
                    coordinates {
                        (2, 0.7)
                        (10,3.5)
                        (20,7.4)
                        (30,11.8)
                        (40,15.3)
                    };
                \addplot+[mark size=3pt]
                    coordinates {
                        (2,0.8)
                        (10,4.2)
                        (20,8.7)
                        (30,13.6)
                        (40,17.9)
                    };
                \legend{Without SGX, Using SGX}
                \end{axis}
            \end{tikzpicture}
    }
    \caption{AlexNet - CIFAR 10}
    \label{exp1c}
\end{subfigure}
~
~
~
\begin{subfigure}[tbh!]{0.4\columnwidth}
    \resizebox{1\columnwidth}{!}
    {
    \begin{tikzpicture}
                \begin{axis}[
                    xlabel={Edge Devices},
                    ylabel={Processing Time (S)},
                    symbolic x coords = {2,10,20,30,40},
                    xticklabel style={anchor= east,rotate=45 },
                    xtick=data,
                    ymax=25,
                    ymin=0,
                    legend pos=north west,
                    ymajorgrids=true,
                    grid style=dashed,
                    legend style={nodes={scale=1, transform shape}},
                    label style={font=\Large},
                    tick label style={font=\Large}
                ]
                \addplot+[mark size=3pt]
                    coordinates {
                        (2,0.7)
                        (10,3.5)
                        (20,7.3)
                        (30,10.9)
                        (40,14.8)
                    };
                \addplot+[mark size=3pt]
                    coordinates {
                        (2,0.8)
                        (10,4.4)
                        (20,8.1)
                        (30,12.7)
                        (40,16.4)
                    };
                \legend{Without SGX, Using SGX}
                \end{axis}
            \end{tikzpicture}
    }
    \caption{LeNet - MNIST}
    \label{exp1d}
\end{subfigure}
\caption{Processing time of secure aggregation process with and without SGX using various machine learning models and datasets considering batch size = 128.}
\label{exp1}
\end{figure}
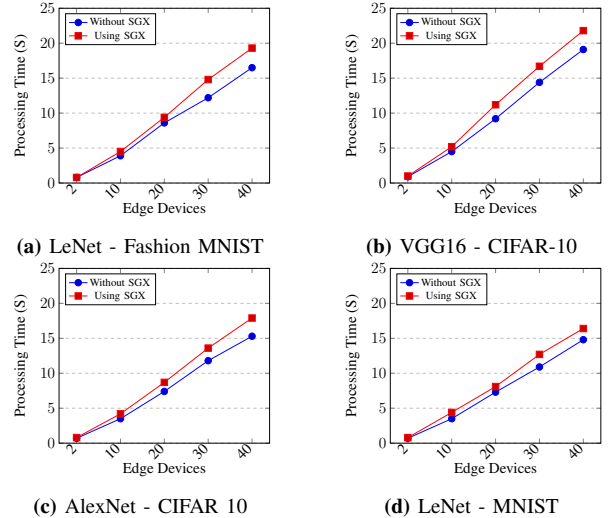

First, we evaluate the performance of our framework for the global model aggregation process(see Fig. \ref{exp1}). This experiment shows the time cost difference when performing the secure aggregation process with enclaves and without enclaves with various numbers of edge devices ranging from 2 to 40. Results show the aggregation times required by LeNet, AlexNet, and VGG16 for a batch size of 128. In Fig. \ref{exp1a} and Fig. \ref{exp1d}, we show the LeNet model with Fashion MNIST and MNIST dataset, respectively. The experimental results show that the LeNet model exhibits a similar trend when used on Fashion MNIST and MNIST datasets. The time is consistently stable when it has 20 edge devices, and the time cost rises a little bit when it reaches 30 edge devices during the aggregation in the TEE. The average additional time cost is 1.2 seconds. Figures \ref{exp1b} and \ref{exp1c} show the results of AlexNet and VGG-16 models with CIFAR-10 dataset. When we perform the VGG-16 model with 40 edge devices, the aggregation process without SGX requires 19.1 seconds. The aggregation process is higher with SGX, which is 21.8 seconds. The required time to aggregate local training models of VGG-16 is the highest due to the involvement of 16 layers. Nevertheless, it is only 2.7 seconds slower than the time cost of aggregation without SGX. Fig. \ref{exp1} shows that the aggregation time cost is 1.3 seconds higher on an average in SGX  due to the paging mechanism and memory limitation of SGX.

\begin{figure}[tbh!]
\centering
\begin{subfigure}[tbh!]{0.4\columnwidth}
    \resizebox{1\columnwidth}{!}
    {
        \begin{tikzpicture}
                    \begin{axis}[
                    xlabel={Batch Size},
                    ylabel={Processing Time (S)},
                    symbolic x coords = {1,8,16,20},
                    xticklabel style={anchor= east,rotate=45 },
                    xtick=data,
                    ymax=15,
                    ymin=0,
                    legend pos=north west,
                    ymajorgrids=true,
                    grid style=dashed,
                    legend style={nodes={scale=1, transform shape}},
                    label style={font=\Large},
                    tick label style={font=\Large}
                ]
                \addplot+[mark size=3pt]
                    coordinates {
                        (1,7.8)
                        (8,8.2)
                        (16,8.7)
                        (20,10.8)
                    };
                \addplot+[mark size=3pt]
                    coordinates {
                        (1,8.3)
                        (8,9.2)
                        (16,9.4)
                        (20,11.3)
                    };
                \legend{Without SGX, Using SGX}
                \end{axis}
            \end{tikzpicture}      
        }
    \caption{LeNet - Fashion MNIST}
    \label{exp2a}
\end{subfigure}
~
~
~
\begin{subfigure}[tbh!]{0.4\columnwidth}
    \resizebox{1\columnwidth}{!}{
    \begin{tikzpicture}
                \begin{axis}[
                    xlabel={Batch Size},
                    ylabel={Processing Time (S)},
                    symbolic x coords = {1,8,16,20},
                    xticklabel style={anchor= east,rotate=45 },
                    xtick=data,
                    ymax=8,
                    ymin=0,
                    legend pos=north west,
                    ymajorgrids=true,
                    grid style=dashed,
                    legend style={nodes={scale=1, transform shape}},
                    label style={font=\Large},
                    tick label style={font=\Large}
                ]
                \addplot+[mark size=3pt]
                    coordinates {
                        (1,1.1047)
                        (8,1.6234)
                        (16,2.2117)
                        (20,3.1469)
                    };
                \addplot+[mark size=3pt]
                    coordinates {
                        (1,1.5234)
                        (8,2.1312)
                        (16,2.9236)
                        (20,4.2643)
                    };
                \legend{Without SGX, Using SGX}
                \end{axis}
            \end{tikzpicture}
        }
    \caption{VGG16 - CIFAR-10}
    \label{exp2b}
\end{subfigure}
~
~
~
\begin{subfigure}[tbh!]{0.4\columnwidth}
    \resizebox{1\columnwidth}{!}
    {
    \begin{tikzpicture}
                \begin{axis}[
                    xlabel={Batch Size},
                    ylabel={Processing Time (Ms)},
                    symbolic x coords = {1,8,16,20},
                    xticklabel style={anchor= east,rotate=45 },
                    xtick=data,
                    ymax=400,
                    ymin=0,
                    legend pos=north west,
                    ymajorgrids=true,
                    grid style=dashed,
                    legend style={nodes={scale=1, transform shape}},
                    label style={font=\Large},
                    tick label style={font=\Large}
                ]
                \addplot+[mark size=3pt]
                    coordinates {
                        (1,98.6)
                        (8,111.4)
                        (16,212.4)
                        (20,275.1)
                    };
                \addplot+[mark size=3pt]
                    coordinates {
                        (1,144.3)
                        (8,152.2)
                        (16,289.7)
                        (20,361.6)
                    };
                \legend{Without SGX, Using SGX}
                \end{axis}
            \end{tikzpicture}
    }
    \caption{AlexNet - CIFAR 10}
    \label{exp2c}
\end{subfigure}
~
~
~
\begin{subfigure}[tbh!]{0.4\columnwidth}
    \resizebox{\columnwidth}{!}
    {
    \begin{tikzpicture}
                \begin{axis}[
                    xlabel={Batch Size},
                    ylabel={Processing Time (Ms)},
                    symbolic x coords = {1,8,16,20},
                    xticklabel style={anchor= east,rotate=45 },
                    xtick=data,
                    ymax=300,
                    ymin=0,
                    legend pos=north west,
                    ymajorgrids=true,
                    grid style=dashed,
                    legend style={nodes={scale=1, transform shape}},
                    label style={font=\Large},
                    tick label style={font=\Large}
                ]
                \addplot+[mark size=3pt]
                    coordinates {
                        (1,51.1)
                        (8,92.7)
                        (16,141.6)
                        (20,184.9)
                    };
                \addplot+[mark size=3pt]
                    coordinates {
                        (1,65.8)
                        (8,120.2)
                        (16,195.9)
                        (20,258.1)
                    };
                \legend{Without SGX, Using SGX}
                \end{axis}
            \end{tikzpicture}
    }
    \caption{LeNet - MNIST}
    \label{exp2d}
\end{subfigure}
\caption{Processing time of training process with and without SGX for different number of batch size using various machine learning models and datasets.}
\label{exp2}
\end{figure}
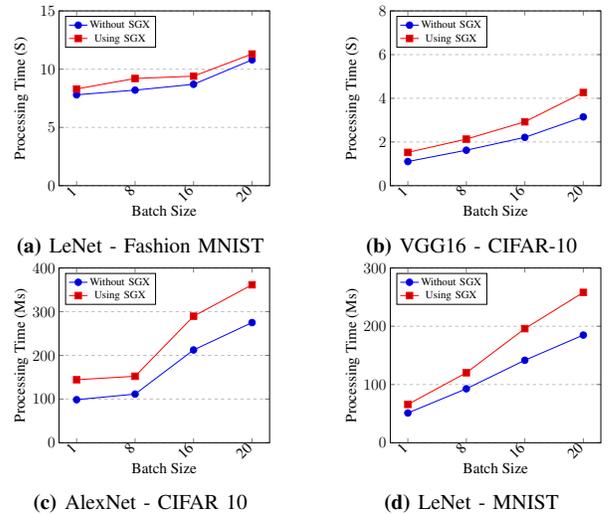

In Fig. \ref{exp2} we test the performance of our framework using different machine learning models and datasets. The experiment is conducted within and outside the enclave with different batch sizes (1, 8, 16, and 20), and the time costs of the training processes are shown in Fig. \ref{exp2a}. The time cost of the LeNet machine learning model with Fashion MNIST datasets running outside the enclave starts from 7.2 seconds for one batch size. The time cost increases linearly to 8.7 seconds for 16 batch size. Fig. \ref{exp2b} shows the time costs of VGG-16 with CIFAR 10 datasets. The time cost is 1.1 seconds for batch size is 1 and 2.2 seconds for batch size 16. The time costs increase slightly, keeping the same linear characteristics when the experiments are performed inside the enclave with the same settings. The LeNet model requires 8.1 seconds to 9.4 seconds, while VGG-16 requires 1.5 seconds to 2.9 seconds. The experiment results show that the time cost increases for both inside and outside enclave training when all the machine learning models use 20 batch sizes. The time costs also increase if the number of images in a batch increases.

%%%%
\begin{figure}[bh]
\centering
\begin{subfigure}[tbh!]{0.4\columnwidth}
    \resizebox{1\columnwidth}{!}{
            \begin{tikzpicture}
                \begin{axis}[
                    xlabel={Number of Nodes},
                    ylabel={Processing Time (Min)},
                    symbolic x coords = {2,10,20,30,40},
                    xticklabel style={anchor= east,rotate=45 },
                    xtick=data,
                    ymax=10,
                    ymin=0,
                    legend pos=north west,
                    ymajorgrids=true,
                    grid style=dashed,
                    legend style={nodes={scale=1, transform shape}},
                    label style={font=\Large},
                    tick label style={font=\Large}
                ]
                \addplot+[mark size=3pt]
                    coordinates {
                        (2,4.49)
                        (10,5.16)
                        (20,5.35)
                        (30,5.42)
                        (40,5.86)
                    };
                \addplot+[mark size=3pt]
                    coordinates {
                        (2,5.78)
                        (10,6.12)
                        (20,6.24)
                        (30,6.56)
                        (40,6.91)
                    };
                \legend{AggregationWithout SGX, Aggregation Using SGX}
                \end{axis}
            \end{tikzpicture}

        }
    \caption{}
    \label{fig:processing_time_tr}
\end{subfigure}
~
~
~
\begin{subfigure}[tbh!]{0.4\columnwidth}
    \resizebox{1\columnwidth}{!}{
        \begin{tikzpicture}
                \begin{axis}[
                    xlabel={Number of Nodes},
                    ylabel={Processing Time (Ms)},
                    symbolic x coords = {5,10,15,20},
                    xticklabel style={anchor= east,rotate=45 },
                    xtick=data,
                    ymax=300,
                    ymin=0,
                    legend pos=north west,
                    ymajorgrids=true,
                    grid style=dashed,
                    legend style={nodes={scale=1, transform shape}},
                    label style={font=\LARGE},
                    tick label style={font=\Large}
                ]
                \addplot+[mark size=3pt]
                    coordinates {
                        (5,98.026)
                        (10,145.24)
                        (15,196.52)
                        (20,236.135)
                    };
                \addplot+[mark size=3pt]
                    coordinates {
                        (5,64.063)
                        (10,119.952)
                        (15,148.845)
                        (20,181.798)
                    };
                \legend{Deploy, Verify}
                \end{axis}
            \end{tikzpicture}
        }
    \caption{}
    \label{fig:processing_time_bc}
\end{subfigure}
\caption{Processing time: (a) for the federated learning process using LeNet model\cite{lecun1998gradient} with Fashion MNIST datasets\cite{xiao2017fashion} with different number of nodes, and (b) for adding the global model to the blockchain with different number of blockchain nodes.}
\label{}
\end{figure}
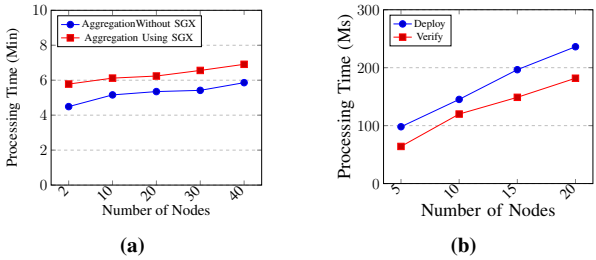

Fig. \ref{fig:processing_time_tr} shows the required time for the FL process with different numbers of edge servers. In this experiment, we performed multiple federated learning processes that use normal aggregation and SGX-based secure aggregation. We used pre-processed the Fashion MNIST dataset and the LeNet machine learning model with 128 batch sizes for the experiment. We consider different number of edge servers ranging from 2 to 40. Results indicate that the time cost increases gradually for the federated learning process with and without SGX-based aggregation. When comparing the results of normal CPU and SGX based approach, the time differs about 70 milliseconds with 10 edge servers and 91 milliseconds with 40 edge servers.

Fig. \ref{fig:processing_time_bc} shows the time required to execute both the verification and deployment of the global model in our blockchain network. In this experiment, the TEE aggregates all the models from the edge server to form a global model. The global model is then verified and deployed in the blockchain network. Our simulation tested the performance using several blockchain nodes ranging from 5 to 20. The deployment phase requires roughly 100 milliseconds (with 5 blockchain nodes) to 230 milliseconds (with 20 blockchain nodes). The verification phase is faster than the deployment phase and requires 60 milliseconds and 180 milliseconds with 5 and 20 blockchain nodes, respectively. The processing times of both phases increase linearly with the increment of blockchain nodes.

%%%%
\begin{table}[h]
\begin{center}
\resizebox{\columnwidth}{!}{
    \begin{tabular}{ |c|c|c|c|c|c| }
        \hline
        
        \textbf{Methodologies} & \textbf{CNN Model} & \textbf{Dataset} & \textbf{Baseline} & \textbf{SGX} & \textbf{\makecell{Accuracy \\ Reduction}}\\
        
        \hline
        Proposed Method & LeNet\cite{lecun1998gradient} & F-MNIST\cite{xiao2017fashion} & 93.2\% & 90.8 & 2.4\%\\
        
        Proposed Method & AlexNet\cite{krizhevsky2017imagenet} & CIFAR-10\cite{krizhevsky2010convolutional} & 73.3\%  & 70.4\% & 2.9\%\\
        
        Proposed Method & VGG-16\cite{simonyan2014very} & CIFAR-10\cite{krizhevsky2010convolutional} & 87.4\% & 84.8\% & 2.6\%\\
        
        Proposed Method & LeNet\cite{lecun1998gradient} & MNIST\cite{deng2012mnist} & 95.7\% & 93.6\% & 2.1\%\\
        
        Myelin\cite{hynes2018efficient} & RESNET-32\cite{szegedy2017inception} & CIFAR-10\cite{krizhevsky2010convolutional} & 89.5\% & 84.4\% & 5.1\%\\
        
        Chiron\cite{hunt2018chiron} & VGG-9\cite{simonyan2014very} & CIFAR-10\cite{krizhevsky2010convolutional} & 88.5\% & 81.1\% & 7.4\%\\
        \hline
    \end{tabular}
    }
    \end{center}
    \caption{Comparison of machine learning model accuracy in federated learning process when using normal CPU and SGX}
    \label{exp5}
\end{table}
%%%%

Table \ref{exp5} shows the results of testing our framework to see the effect when we apply the machine learning model in a federated way inside the enclave and standard CPU. In this experiment, all the datasets have 28x28 pixels and 128 batch size. We ran the experiment with 50 training iterations. The experimental results show that the differences in the accuracy of the proposed methodology and two benchmark methods proposed in \cite{hynes2018efficient} and \cite{hunt2018chiron}. Initially, we record the accuracies of our proposed method with and without SGX. The accuracies of the aforementioned methods are obtained by applying various CNN models on different datasets. According to the results, the accuracies are reduced by 2.2\% to 2.9\% when SGX is used. Later, we measure the accuracies of Myelin\cite{hynes2018efficient} and Chiron\cite{hunt2018chiron} with and without SGX. Results show that accuracies of the Myelin and Chiron are lower than our proposed method. Moreover, the accuracies of Myelin and Chiron are reduced around 5.1\% and 7.4\% with SGX, respectively. Hence, our method has better accuracy compared to Myelin and Chiron.

\subsection{Discussion}
In this section, we summarize the performance of our proposed method. As discussed in Section \ref{sec:results}, we conducted a series of experiments to evaluate the efficacy of our proposed method. Based on the empirical results, the following conclusions can be drawn.

\begin{itemize}

    \item \textbf{Privacy of Local Dataset}:
    Federated learning allows computational parties to collaboratively learn a shared model while preserving all training data locally, separating the machine learning process from the storage of data in the central server. The method is unlike traditional centralized machine learning where local datasets are stored in one central server. Therefore, federated learning can ensure the privacy of the client's sensitive data.

    \item \textbf{Privacy of Local Training Model}:
    In our framework, the local training model is encrypted using a shared key before it is sent to the blockchain node. The shared key is established using a secure key-exchange protocol. Later, the local training model will be decrypted inside the enclave for secure aggregation. As the local model is encrypted, model inversion attacks \cite{fredrikson2015model}, and parameter stealing \cite{8418595} cannot be performed on a local model by an adversary.
    
    \item \textbf{TEE-based Secure Aggregation}:
    In federated learning, aggregation is typically performed on a normal server. Several researchers \cite{wei2020federated,9253545} have proposed a differential privacy (DP) method to secure the model during the aggregation process. However, DP will significantly reduce the accuracy of the global model. Table. \ref{exp5} shows that the use of secure TEE-based aggregation can overcome this problem while maintaining the privacy of the model. As the aggregation is performed in the TEE, adversaries cannot tamper with or steal the model parameters during the aggregation process. As blockchain technology is being used with emerging technologies, such as drones\cite{rahman2021blockchain,8795473,rahman2021blockchain1}, the proposed blockchain and TEE-based model aggregation in FL would enhance the trust in applications where drones are used as edge intelligence in FL \cite{9635590,9670460}. 
    
    \item \textbf{Resilience of the Global Model}:
    Blockchain is a decentralized technology that can maintain data integrity by means of an extensive network that can withstand security breaches from untrusted parties. In the proposed framework, we use blockchain to store the global model after the aggregation process in the TEE. This decentralization makes it almost impossible for an adversary to compromise the network. Moreover, model updates are protected by digital signatures and hashes. Hence, the adversary cannot tamper with or contaminate the global model since this will change the hash value.
    
    \item \textbf{Model Performance}:
    Although our proposed method can ensure the privacy of the model and the security of the aggregation process, performance is still a crucial metric for measuring the quality of the framework. The experimental results show that the performance of the proposed framework is better than that of the baseline model. Our proposed framework is different from \cite{tramer2018slalom}, \cite{juvekar2018gazelle}, \cite{hynes2018efficient}, and \cite{hunt2018chiron} where the whole training process occurs inside the enclave for a single deep-learning model. On the other hand, our framework uses a federated learning setup, and only the aggregation is performed inside the enclave. We also examine the reduction of our model’s accuracy when we leverage TEE. Our proposed method has only up to 3\% accuracy reduction compared to Myelin \cite{hynes2018efficient}, and Chiron \cite{hunt2018chiron} that have more than 5\% and 7\% accuracy reduction, respectively. In other words, our proposed framework achieves a good balance between privacy and model performance.

\end{itemize}

\section{Conclusion}\label{sec:con}
In this paper, a blockchain and Trusted Execution Environment (TEE) enabled Federated Learning (FL) framework is proposed for IoT. The main objective of this framework is to ensure the trustworthy aggregation of local models to obtain a global model. The aggregation is done within the blockchain network. The proposed framework leverages the Intel Software Guard Extension (SGX)-based Trusted Execution Environment (TEE) to ensure secure aggregation where each blockchain node executes the aggregation task. In this framework, each blockchain node is equipped with an SGX-enabled processor that securely generates a global model to ensure trustworthiness. Later, the global model is verified by the blockchain network via a consensus mechanism before it is added to the blockchain, thereby maintaining tamperproof storage. Users of the global model can access it and verify its integrity only through the blockchain network. We use different Convolutional Neural Network (CNN) based algorithms with several benchmark datasets to generate local models and aggregate them under FL settings. We conducted several experiments that show that our proposed framework’s processing time is almost similar to that of the original FL model. In addition, our framework has only around 2\% less accuracy compared to the original FL model. It is essential to mention that this framework has leveraged a hash-based consensus mechanism to ensure the model’s integrity. In the future, we intend to develop an efficient consensus mechanism for the proposed TEE and blockchain-based FL framework in order to make it more practical. In this paper, we assume that all participants perform homogeneous tasks and use same approach to generate their respective local models. Each participant uses their own private dataset and the federated learning architecture to obtain a global model. However, we plan to extend our current work in the future to support heterogeneous tasks in Blockchain-based federated learning with TEE based secure aggregation.

\section*{Acknowledgement}
This work is supported by the Australian Research Council Discovery Project (DP210102761).

\bibliographystyle{IEEEtran}
\bibliography{References}

\end{document}